\documentclass[twocolumn,showpacs,aps]{revtex4}
\usepackage{graphicx}
\usepackage{graphics}
\usepackage{amsmath}

\begin{document}

\title{Universal energy distribution for interfaces in a random field
environment}
\author{Andrei A. Fedorenko and Semjon Stepanow }
\affiliation{Martin-Luther-Universit\"{a}t Halle, Fachbereich Physik, D-06099\\
Halle, Germany}
\date{\today }

\begin{abstract}
We study the energy distribution function $\rho (E)$ for interfaces in a
random field environment at zero temperature by summing the leading terms in
the perturbation expansion of $\rho (E)$ in powers of the disorder strength,
and by taking into account the non perturbational effects of the disorder
using the functional renormalization group. We have found that the average
and the variance of the energy for one-dimensional interface of length $L$
behave as, $\langle E\rangle _{R}\propto L\ln L$, $\Delta E_{R}\propto L$,
while the distribution function of the energy tends for large $L$ to the
Gumbel distribution of the extreme value statistics.
\end{abstract}

\maketitle

The concept of energy landscapes is of current interest in different topics
such as structural glasses, spin glasses, proteins, flux lines etc. [%
\onlinecite{stillinger}-\onlinecite{kimetal91}]. The existence of metastable
states is crucial for the complex behavior in these systems. The domain wall
counterpart of the random field Ising model \cite{rfim} provides an example
of the problem which possesses complex properties, and can be quantitatively
treated using the well established analytical methods such as the functional
renormalization group (FRG) method \cite{fisher86} and the method of replica
symmetry breaking (RSB) \cite{mezard-parisi90}. A significant progress has
been achieved in recent years in understanding the behavior of interfaces in
disordered media at equilibrium \cite{fisher86} and the driven interfaces at
the depinning transition [\onlinecite{nstl92}-\onlinecite{chauve01}]. It is
expected that in equilibrium or below the depinning transition there are
many metastable states constituting the energy landscape. This makes the
interface problem a natural candidate to study the concepts of energy
landscapes. For recent theoretical and numerical studies of the related
systems under the perspective of the energy landscape see Refs.~[%
\onlinecite{balents96}-\onlinecite{balents-ledoussal02}]. In this Letter we
present the results of the study of the distribution function (DF) of the
energy $\rho (E)$ for an interface in a random field environment at
equilibrium. The main result of this Letter is that for large interfaces $%
\rho (E)$ is a universal function which coincides with the Gumbel
distribution of the extreme value statistics. The dynamic formalism we use
here can be applied to the study of the dynamic aspects of energy landscapes
such as the relaxation of the energy, the two times energy correlation
functions etc.

The interface motion in a disordered medium at $T=0$ is described by the
equation
\begin{equation}
\mu ^{-1}\frac{\partial z(x,t)}{\partial t}=\gamma \nabla ^{2}z+F+g(x,z),
\label{e1}
\end{equation}%
where $\mu $\ is the mobility, $\gamma $ is the stiffness constant, and $F$
is the driving force density. The quenched random force $g(x,z)$ is assumed
to be Gaussian distributed with the zero mean and the second cumulant $%
\left\langle g(x,z)g(x^{\prime },z^{\prime })\right\rangle =\delta
^{d}(x-x^{\prime })\Delta (z-z^{\prime })$, where $\Delta (z)$ is a short
ranged function with the width $a$, and $d$ is the interface dimension.

It is well-known that the Langevin equation (\ref{e1}) can be reformulated
in terms of the Fokker-Planck equation for the conditional probability
density $P(z(x),t;z^{0}(x),t^{0})$ to have the profile $z(x)$ at time $t$ by
having the profile $z^{0}(x)$ at time $t^{0}$. This Fokker-Planck equation
can be written as an integral equation, which, for an interface of a finite
length $L$, reads
\begin{eqnarray}
P(z,t;z^{0},t^{0}) &=&P_{0}(z,t;z^{0},t^{0})  \notag \\
&-&\mu \int_{t_{0}}^{t}dt^{\prime }\int \mathcal{D}z^{\prime
}P_{0}(z,t;z^{\prime },t^{\prime })  \notag \\
&\times &\sum_{k^{\prime }}\partial _{z_{k^{\prime }}^{\prime }}g_{k^{\prime
}}(z^{\prime })P(z^{\prime },t^{\prime };z^{0},t^{0}),  \label{e3}
\end{eqnarray}%
where $z_{k}=\int d^{d}xz(x)\exp (-ikx)$ and $g_{k}(z)=\int d^{d}x\exp
(-ikx)g(x,z)$ ($k=(k_{1},...,k_{d})$, $k_{i}=2\pi j_{i}/L$, $j_{i}=0,\pm
1,...$) are the Fourier transforms of the interface height and the quenched
force, respectively. $\int \mathcal{D}z$ in (\ref{e3}) stays for
integrations over the modes $z\equiv \{z_{k}\}$. The bare conditional
probability for non zero modes reads
\begin{equation}
P_{0}(z,t;z^{0},t^{0})=\prod\limits_{k}\delta (z_{k}-z_{k}^{0}\exp (-\gamma
\mu k^{2}(t-t^{0}))).  \label{e4}
\end{equation}%
Analogously to the case of one Brownian particle \cite{wiegel} the formal
solution of Eq.~(\ref{e3}) can be represented as a path integral.

The probability DF of the energy $E(z)=E_{\mathrm{el}}(z)+E_{\mathrm{dis}}
(z)=\int d^{d}x(\frac{\gamma }{2}(\nabla z)^{2}-\int_{0}^{z(x)}g(x,z))$ can
be calculated using the conditional probability density $P(z(x),t;0,0)$ as
follows
\begin{equation}
\rho (E(t))=\int \mathcal{D}z(x)\delta (E-E(z))P(z(x),t;0,0).  \label{e8}
\end{equation}%
It is convenient instead of $\rho (E(t))$ to consider its Fourier transform $%
\hat{\rho}(s)$ which is obtained from Eq.~(\ref{e8}) as
\begin{equation}
\hat{\rho}(s)=\sum_{n=0}^{\infty }\frac{(-is)^{n}}{n!}%
\sum_{m=0}^{n}C_{n}^{m}\left\langle E_{\mathrm{el}}^{m}(t)E_{\mathrm{dis}%
}^{n-m}(t)\right\rangle .  \label{e9}
\end{equation}%
In this Letter we will restrict ourselves to the study of the energy DF in
the steady state, i.e. for $t\rightarrow \infty $. In this limit $%
\left\langle E_{\mathrm{el}}^{m}E_{\mathrm{dis}}^{n-m}\right\rangle $ is
related to the static equilibrium correlation function
\begin{eqnarray}
\left\langle z(x_{1})z(x_{2})\right\rangle &=&\lim_{t\rightarrow \infty
}\int \mathcal{D}z(x)z(x_{1})z(x_{2})  \notag \\
&&\hspace*{-13mm}\times P(z(x),t;0,0)=\int_{k}\frac{\Delta (0)}{(\gamma
k^{2})^{2}}e^{ik(x_{1}-x_{2})}.  \label{e10}
\end{eqnarray}

\begin{figure}[tbph]
\includegraphics[clip,width=2.5in]{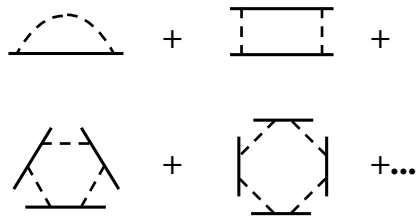}
\caption{The loop expansion of connected diagrams contributing to the energy
distribution function.}
\label{fig1}
\end{figure}

Let us for simplicity elucidate the computation of the \textit{nth} moment
of the elastic energy%
\begin{equation}
\left\langle E_{\mathrm{el}}^{n}\right\rangle =(\frac{\gamma }{2})^{n}\int
\mathcal{D}z\int_{k_{1}}k_{1}^{2}\left\vert z_{k_{1}}\right\vert
^{2}...\int_{k_{n}}k_{n}^{2}\left\vert z_{k_{n}}\right\vert ^{2}P(z,t;0,0),
\label{e11}
\end{equation}%
where we expressed $E_{\mathrm{el}}$ through the Fourier components of the
interface height $z(x)$. For an interface of a finite size $L$ the integral $%
\int_{k}$ means $L^{-d}\sum\nolimits_{k}$. To compute (\ref{e11}) to the
lowest order in disorder strength we iterate Eq.~(\ref{e3}) $2n$ times and
insert it into (\ref{e11}). Expecting that the steady state does not depend
on the initial interface configuration we have taken the latter in (\ref{e10}%
) and (\ref{e11}) to be flat at $t_{0}=0$. The average over the random
forces, which is carried out by using the Wick theorem, yields connected and
disconnected expressions. The connected expression contains only one
integration over $k$, while the number of integrations over $k$ in a
disconnected expression is equal to the number of connected parts in that
expression. Let us consider the calculation of the connected part of $%
\left\langle E_{\mathrm{el}}^{n}\right\rangle $. As a result of integrations
by parts in (\ref{e11}) with $P(z,t;0,0)$\ being iterated $2n$ times the $2n$
derivatives with respect to $z_{k_{i}^{\prime }}^{\prime }$ (see Eq.~(\ref%
{e3})) will act on $z_{k_{i}}$ in (\ref{e11}). This has the consequence that
pairs of $2n$ momenta $k_{1}^{\prime }$, ..., $k_{2n}^{\prime }$ associated
with the right-hand side of (\ref{e3}) (being iterated) become equal
consecutively to one of $k_{1}$, ..., $k_{n}$ in (\ref{e11}). The number of
such possibilities is $(2n)!$. The rid of $2n$ ordered time integrations in $%
P(z,t;0,0)$ gives the factor $1/(2n)!$ The number of possibilities to get a
connected loop diagram shown in Fig.~\ref{fig1} with $n$ continuous lines is
$2^{n-1}(n-1)!$. Integrations over $x_{1}$, ..., $x_{2n-1}$ arising from the
above expression of $g_{k}(z)$ provides that the momenta of the modes being
connected by a dashed line, which is associated with the disorder
correlator, become equal. The integration over $x_{2n}$ gives the factor $%
L^{d}$. The intermediate $z_{k}^{\prime }$ are zero for flat initial
interface configuration due to delta functions in (\ref{e4}). As a result
the arguments of disorder correlators $\Delta (z)$ become zero. Collecting
all combinatorial factors and taking the limit $t\rightarrow \infty $ we
find the following expression of the connected part of $\left\langle E_{%
\mathrm{el}}^{n}\right\rangle $%
\begin{equation}
\frac{1}{n!}\left\langle E_{\mathrm{el}}^{n}\right\rangle _{c}=\frac{1}{2n}%
\Delta (0)^{n}\gamma ^{-n}L^{d}\int_{k}\frac{1}{(k^{2})^{n}}.  \label{e12}
\end{equation}%
The computation of $\left\langle E_{\mathrm{el}}^{m}E_{\mathrm{dis}%
}^{n-m}\right\rangle $ to the same order is similar and gives $\left\langle
E_{\mathrm{el}}^{m}E_{\mathrm{dis}}^{n-m}\right\rangle
=\frac12(-2)^{n-m}(n-1)!\Delta (0)^{n}\gamma ^{-n}L^{d}\int_{k}1/(k^{2})^{n}$%
. The use of the latter yields $\frac{1}{n!}\left\langle E^{n}\right\rangle
_{c}$ which is obtained from (\ref{e12}) multiplied by the factor $(-1)^{n}$%
. The expression $\frac{1}{n!}\left\langle E^{n}\right\rangle _{c}$ is
associated with the loop diagram consisting of \thinspace $n$ continuous
lines (see Fig.~\ref{fig1}). The factor $2n$ is the symmetry number of the
diagram.
The straightforward analysis gives that the expansion (\ref{e9}) can be
represented as a series of loop diagrams. The use of the connectivity
theorem enables us to write the Fourier transform of the energy DF (\ref{e9}%
) as exponential of the series of connected loop diagrams shown in Fig.~\ref%
{fig1}
\begin{equation}
\hat{\rho}(s)=\exp \left( -\frac{1}{2}L^{d}\int_{k}\ln \left( 1-\frac{%
is\Delta (0)}{\gamma k^{2}}\right) \right) .  \label{e13a}
\end{equation}%
Notice that $\hat{\rho}(s)$ given by the diagram series in Fig.~\ref{fig1}
is closely related to the loop expansion of the effective potential in
Quantum Field Theory studied in \cite{coleman-weinberg}. Replacing the
integral in (\ref{e13a}) by the sum according to $L\int_{k}f(k)\rightarrow
\sum_{j=-\infty }^{\infty }f(2\pi j/L)$ we find in $d=1$
\begin{equation}
\hat{\rho}(s)=\prod\limits_{j=1}^{\infty }\left( 1+isE_{0}/j^{2}\right)
^{-1}=\frac{\pi \sqrt{isE_{0}}}{\sinh (\pi \sqrt{isE_{0}})},  \label{e13}
\end{equation}%
where $E_{0}=-\Delta (0)L^{2}/(4\pi ^{2}\gamma)$ is the characteristic
energy for an interface with the perturbational roughness $w\propto L^{3/2}$%
, which follows from $w\propto L^{(4-d)/2}$ for $d=1$.

Eq.~(\ref{e13}) has only simple poles $s=ij^{2}/E_{0}$ in the lower
half-plane, so that the inverse Fourier transformation of (\ref{e13}) can be
easily performed as a sum over all poles by using the Jordan's lemma. As a
result we obtain the DF as $\rho (E)=|E_{0}|^{-1}f\left( E/E_{0}\right) $, $%
E<0$ , where
\begin{equation}
f(x)=2\sum\limits_{j=1}^{\infty }(-1)^{(j+1)}j^{2}e^{-xj^{2}}.  \label{e17}
\end{equation}%
The derivation of $\rho (E)$ shows that (\ref{e17}) describes also the DF of
the elastic energy ($x=E/|E_{0}|>0$) and the disorder energy ($x=E/2E_{0}>0$%
). Eq.~(\ref{e17}) coincides with the dimensionless width DF for the
one-dimensional random-walk interface studied in Ref.~\cite{oerding94}.
Using the method of stationary phase it was shown in \cite{oerding94} that (%
\ref{e17}) can be well approximated for small $x$ by $f(x)\approx \sqrt{\pi
/x^{5}}(\pi ^{2}/2-x)e^{-\pi ^{2}/4x}$. Using (\ref{e17}) we have computed
the average energy, $\langle E\rangle =\pi ^{2}E_{0}/6$, and the variance $%
\Delta E=(\left\langle E^{2}\right\rangle -\left\langle E\right\rangle
^{2})^{1/2}=\pi ^{2}|E_{0}|/(3\sqrt{10})$. Eq.~(\ref{e17}) is the exact
perturbational result generalizing the result established by Efetov and
Larkin \cite{efetov-larkin77} for the height-height correlation functions (%
\ref{e10}), which can be readily proved using supersymmetry \cite{parisi}.
Nevertheless, both (\ref{e10}) and (\ref{e17}) are wrong due to the fact
that (\ref{e10}) gives the value $(4-d)/2$ for the roughness exponent
instead of the correct Imry-Ma \cite{imry-ma} value $\zeta =(4-d)/3$.

\begin{figure}[tbph]
\includegraphics[clip,width=2.8in]{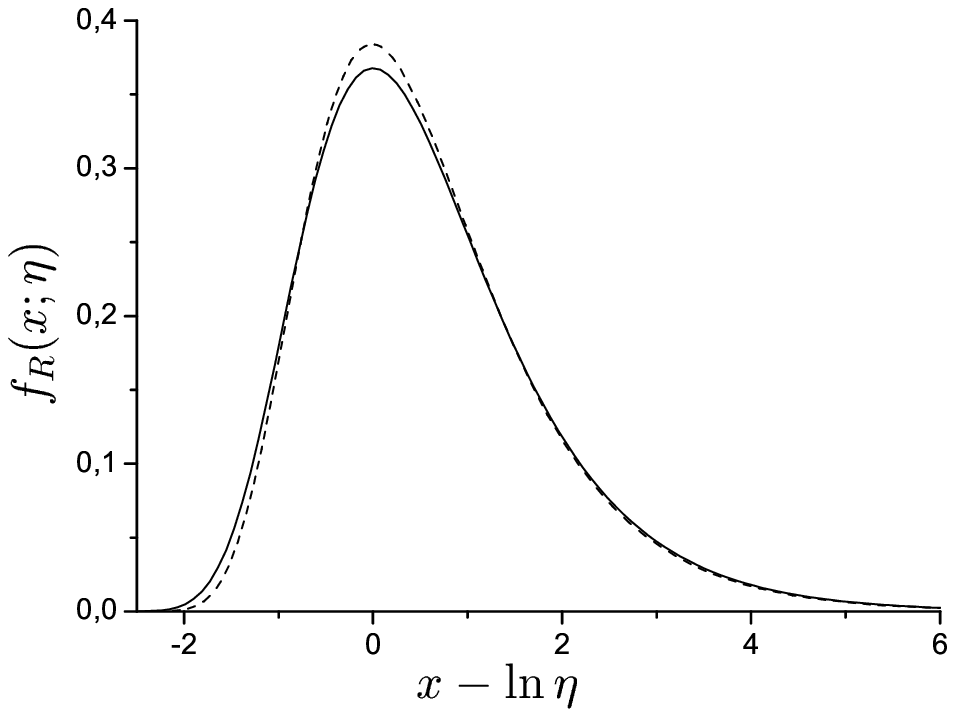}
\caption{The renormalized distribution of the energy for a line in a random
field environment. Dashed line: $L/L_{c}=10^{2}$; solid line: the Gumbel
distribution.}
\label{fig2}
\end{figure}

We now will take into account the effect of the renormalization on the
energy DF using the results of the FRG \cite{fisher86}. According to \cite%
{fisher86} the renormalized disorder correlator $\Delta _{R}(0)$ acquires
the scale dependence $l^{2\zeta -\varepsilon }$. Taking into account the
latter in (\ref{e10}) gives the correct value of the roughness exponent $%
\zeta $. To enable a crossover to the perturbational regime at small length
scales we use the ansatz
\begin{equation}
\Delta _{R}(0)=\Delta (0)\left[ 1+\left( k/k_{c}\right) ^{2\zeta
-\varepsilon }\right] ^{-1},  \label{e21}
\end{equation}%
where $\varepsilon =4-d$, and the wave-vector $k_{c}=2\pi /L_{c}$ is
associated with the Larkin length $L_{c}\simeq \left( \gamma
^{2}a^{2}/\Delta (0)\right) ^{1/\varepsilon }$. The ansatz (\ref{e21})
describes the scale dependence of $\Delta _{R}(0)$ at the cusped fixed-point
solution of the disorder correlator, $\Delta _{R}(0)\simeq \Delta (0)\left[
k/k_{c}\right] ^{\varepsilon -2\zeta }$ for $k\ll k_{c}$, and describes the
crossover to the perturbational regime, $\Delta _{R}(0)\simeq \Delta (0)$
for $k\gg k_{c}$. Using the renormalized $\Delta _{R}(0)$ in Eq.~(\ref{e13a}%
) we obtain the Fourier transform of the renormalized distribution of the
energy in $d=1$ as
\begin{equation}
\widehat{\rho }_{R}(s)=\prod\limits_{j=1}^{\infty }\left( 1+\frac{is\tilde{E}%
_{0}}{j(1+j/\eta )}\right) ^{-1},  \label{e21a}
\end{equation}%
where $\eta =L/L_{c}$ and $\tilde{E}_{0}=-\Delta (0)L_{c}L/(4\pi ^{2}\gamma
) $. Carrying out the inverse Fourier transformation of (\ref{e21a}) we
obtain $\rho _{R}(E)=|\tilde{E}_{0}|^{-1}f_{R}(E/\tilde{E}_{0};\eta )$,
where the function $f_{R}(x;\eta )$ is given by
\begin{equation}
f_{R}(x;\eta )=\sum\limits_{j=1}^{\infty }(-1)^{j+1}\frac{\Gamma (j+\eta
+1)(1+2j/\eta )}{\Gamma (\eta +1)\Gamma (j)}e^{-j(1+j/\eta )x}.  \label{e23}
\end{equation}%
For string lengths much shorter than the Larkin length, $\eta \ll 1$, the DF
(\ref{e23}) pass over to the perturbational result (\ref{e17}). Similar to
the height-height correlation function at equilibrium we expect that (\ref%
{e23}), which is the result of the renormalization of (\ref{e17}) to order $%
\varepsilon $, is exact. Eq.~(\ref{e23}) applies to order $\varepsilon $ at
the depinning transition too with the difference that in this case there are
corrections to (\ref{e23}) of order $\varepsilon ^{2}$. However, we expect
that the latter will be small as it is the case for corrections of order $%
\varepsilon ^{2}$ to the interface width distribution at the depinning
transition \cite{rosso03}. The average energy $\langle E\rangle _{R}$
derived from Eq.~(\ref{e21a}) is
\begin{eqnarray}
\langle E\rangle _{R} &=&\tilde{E}_{0}\sum_{j=1}^{\infty }\frac{1}{%
j(1+j/\eta )}=\left[ \Psi (\eta +1)+C\right] \tilde{E}_{0}  \notag \\
&\simeq &\left[ \ln \eta +C\right] \tilde{E}_{0}+O\left( 1/\eta \right)
\propto L\ln L,  \label{new1}
\end{eqnarray}%
where $\Psi (x)$ is the digamma function and $C$ $=0.5772...$ is the Euler's
constant. The energy fluctuation $\Delta E_{R}$ obtained from (\ref{e21a})
reads
\begin{eqnarray}
\Delta E_{R} &=&|\tilde{E}_{0}|\left[ \frac{\pi ^{2}}{6}+\Psi ^{\prime
}(\eta +1)-2(C+\Psi (\eta +1))/\eta \right] ^{1/2}  \notag \\
&\simeq &\frac{\pi }{\sqrt{6}}|\tilde{E}_{0}|+O\left( \ln \eta /\eta \right)
\propto L.  \label{e24}
\end{eqnarray}%
The result, $\Delta E_{R}\propto L$, agrees with the estimate of the energy
by using dimensionality arguments with correct roughness exponent $\zeta $.
Notice that due to the logarithmic term in (\ref{new1}) $\langle E\rangle
_{R}$ and $\Delta E_{R}$ scale in different way, and the relative
fluctuation, $\Delta E_{R}/\langle E\rangle _{R}$, disappears as $1/\ln L$
for large $L$, which is in contrast to $1/\sqrt{L}$ behavior for a Gaussian
distribution. The latter reflects the relevance of fluctuations over all
length scales. The higher moments of the energy distribution (\ref{e23})
scale as $\langle (E-\langle E\rangle )^{n}\rangle \propto \Delta E_{R}^{n}$%
. Fig.~\ref{fig2} shows $f_{R}(x;\eta )$, which is given by Eq.~(\ref{e23}),
as a function of $x-\ln \eta $.

We now will consider the asymptotic behavior of (\ref{e23}) in the limit of
long lines, $L\gg L_{c}$. Changing $x$ in favor of $x-\ln \eta \equiv y$ and
taking the limit $\eta \rightarrow \infty $ we calculate the sum over $j$ in
(\ref{e23}) and arrive at
\begin{equation}
f_{R}(y)=\mathcal{P}(y)\equiv \exp (-y-\exp (-y)),  \label{e22}
\end{equation}%
which is nothing but\ the Gumbel distribution of the extreme value
statistics \cite{gumbel58}. The universality of $f_{R}(y)$\ is due to the
universal character of fluctuations on large scales, which are described by
the fixed-point solution of the FRG \cite{fisher86}. Notice that the
expectation value of $y$ calculated with (\ref{e22}) gives the Euler's
constant $C$ which is in consistence with Eq.~(\ref{new1}) of the average
energy. We have checked that the limit of the distribution $f_{R}(x;\eta )$
for $\eta \rightarrow \infty $ is insensitive to the details of the
renormalization at scales smaller than the Larkin scale.

The Gumbel distribution is one of the three possible limit distributions in
the extreme value statistics \cite{gumbel58}, which concern the distribution
of the maximum $M_{n}=\mathrm{max}\{\xi _{1},...,\xi _{n}\}$ of the set of
identically distributed random variables $\xi _{i}$ ($i=1,2,...,n$). The
asymptotic distribution $P_{n}(x)$ for $M_{n}$ in limit $n\rightarrow \infty
$ does not depend on details of the distribution of $\xi _{i}$ and under
fulfilling some conditions \cite{gumbel58} has the form $P_{n}(x)\simeq
\mathcal{P}(x-\ln n)$ where $\mathcal{P}(y)$ is given by Eq.~(\ref{e22}).
The combination $y=x-\ln n$, where $n$ is the number of random variables
guarantees that the distribution remains invariant for $n\rightarrow \infty $%
.

Vinokur \textit{et al}. \cite{vinokur96} have used the Gumbel distribution
to describe in a phenomenological way the energy barriers distribution for a
flux line in a random environment. The creep motion of the flux line in the
limit of small driving force $F$ and low temperature is controlled by
thermally activated jumps. The thermally activated advance of the flux
segment of length $L$ is controlled by the global barrier $U=\mathrm{max}%
\{U_{1},...,U_{n}\}$, where $U_{i}$ is the barrier for the subsegment $i$ of
length $L_{c}$ with the number of subsegments $n=L/L_{c}$. It was suggested
in \cite{vinokur96} that the probability distribution for a given segment $L$
is $\mathcal{P}(U/U_{c}-\ln (L/L_{c}))$, where $U_{c}\simeq \gamma
a^{2}L_{c}^{d-2}$ is the minimum average barrier between neighboring
metastable positions of a pinned segment $L_{c}$, so that the typical
barrier of a segment of length $L$ scales then as $U\propto U_{c}\ln
(L/L_{c})$. Bouchaud and M\'{e}zard \cite{bouchaud97-2} showed that the
Gumbel distribution describes the energy distribution in a class of random
energy models possessing the one-step RSB. The Gumbel and related
distributions were used in \cite{bramwell}, to describe universal
fluctuations in correlated systems. It was shown in \cite{antal01} that the
Gumbel distribution appears in systems with $1/f$ power specta.

The relation of the energy distribution with the extreme value statistics
can be illustrated as follows. It is known from the treatment of the problem
in the framework of the replica variational approach \cite{mezard-parisi90}
that the system under consideration demonstrates RSB \cite{note1}. The RSB
is related to the existence of many metastable states and to the lost of the
ergodicity of the system. On the contrary, the perturbational result (\ref%
{e17}) corresponds to the average over states which belong to different
nonergodic subsystems. The metastable states, which correspond to the minima
of the total energy of the interface, are rare states among other states of
the energy landscape, and has to reflect the statistics of these extreme
states \cite{bouchaud97-2}. It was shown in \cite{drossel-kardar95,balents96}
that energy minima behave in the same way as barriers, which as suggested in
\cite{vinokur96} are distributed in accordance with the Gumbel distribution.

In conclusion, we have studied the distribution function of the
elastic, disorder, and the total energies of an interface in a random field
environment by summing the leading terms of the perturbation expansion in
powers of the disorder. The nonperturbational effects of the disorder are
taken into account using the FRG method. We have found that the average and
the fluctuation of the energy for one-dimensional interfaces behave as, $%
\langle E\rangle _{R}\propto L\ln L$, $\Delta E_{R}\propto L$, while the
energy DF tends for large $L$ to a universal function which coincides with
the Gumbel distribution.

A support from the Deutsche Forschungsgemeinschaft (SFB 418) is gratefully
acknowledged.


\begin{thebibliography}{99}
\bibitem{note1} The RSB solution \cite{mezard-parisi90} gives the same form
of the height-height correlation function and the same roughness exponent $%
\zeta $ as that predicted by the FRG, and consequently the same energy
distribution function.
\end{thebibliography}
\end{document}